\documentclass[traditabstract,a4paper]{aa}
\usepackage{txfonts} 

\usepackage{color}

\usepackage{natbib}
\bibpunct{(}{)}{;}{a}{}{,} 

\usepackage{graphicx}
\usepackage{fixltx2e}

\usepackage{longtable}
\usepackage{multirow}
\usepackage{overpic}
\usepackage{array}

\usepackage[table,usenames,dvipsnames]{xcolor}
\definecolor{RoyalPurple}{cmyk}{0.75,0.9,0,0.1}
\definecolor{MyBlue}{rgb}{0.0025,0.1125,0.2}
\usepackage[breaklinks, colorlinks, citecolor=blue, linkcolor=MyBlue, urlcolor=RoyalPurple, 
            colorlinks=true, linkcolor=blue, debug, baseurl=' ']{hyperref}

\graphicspath{{Figures/}}

\def\Planck{\textit{Planck}}

\begin{document}

\title{Impact of particles on the \Planck\ HFI detectors:
Ground-based measurements and physical interpretation}

\author{A.~Catalano  \inst{1}
\and
P.~Ade \inst{2}
\and
Y.~Atik  \inst{3}
\and
A.~Benoit  \inst{4}
\and
E.~Br\'eele  \inst{5}
\and
J.J.~Bock \inst{6}$^,$\inst{7}
\and
P.~Camus  \inst{4}
\and
M.~Chabot  \inst{8}
\and
M.~Charra  \inst{3}
\and
B.P.~Crill \inst{7}
\and    
N.~Coron  \inst{3}
\and
A.~Coulais  \inst{9}
\and  
F.-X.~D\'esert  \inst{10}
\and 
L.~Fauvet  \inst{11}
\and 
Y.~Giraud-H\'eraud  \inst{5}
\and 
O.~Guillaudin  \inst{1}
\and
W.~Holmes \inst{7}
\and
W. C.~Jones \inst{12}
\and
J.-M.~Lamarre  \inst{9}
\and      
J.~Mac\'{\i}as-P\'erez  \inst{1}
\and
M.~Martinez  \inst{3}
\and    
A.~Miniussi  \inst{3}
\and 
A.~Monfardini  \inst{4}
\and    
F.~Pajot  \inst{3}
\and
G.~Patanchon  \inst{5}
\and
A.~Pelissier  \inst{1}
\and
M.~Piat  \inst{5}
\and 
J.-L.~Puget  \inst{3}
\and
C.~Renault  \inst{1}
\and
C.~Rosset  \inst{5}
\and
D.~Santos  \inst{1}
\and
A.~Sauv\'e  \inst{13}
\and
L.D.~Spencer \inst{2}
\and
R.~Sudiwala \inst{2}
 }

\offprints{A. Catalano - catalano@lpsc.in2p3.fr} 

\institute{Laboratoire de Physique Subatomique et de Cosmologie, 
  CNRS/IN2P3, Universit\'e Joseph Fourier Grenoble I,
  Institut National Polytechnique de Grenoble, 
  53 rue des Martyrs, 38026 Grenoble Cedex, France 
\and School of Physics and Astronomy, Cardiff University,
  Queens Buildings, The Parade, Cardiff, CF24 3AA, UK
\and Institut d'Astrophysique Spatiale, CNRS (UMR8617)
  Universit\'e Paris-Sud 11, B\^atiment 121, Orsay, France
\and Institut N\'eel, CNRS, Universit\'e Joseph Fourier Grenoble I,
  25 rue des Martyrs, Grenoble, France 
\and Astroparticule et Cosmologie, CNRS
  (UMR 7164), Universit\'e Denis Diderot Paris 7, B\^atiment
  Condorcet, 10 rue A. Domon et Leonie Duquet, Paris, France
\and California Institute of Technology, Pasadena, California, USA
\and Jet Propulsion Laboratory, California Institute of Technology,
  4800 Oak Grove Drive, Pasadena, California, USA
\and IPN: Institut de Physique Nucl\'eaire, CNRS/IN2P3, Université Paris-Sud 11, 91406 Orsay cedex, France
\and LERMA, CNRS, Observatoire de Paris, 61 avenue de 
  l'Observatoire, Paris, France
\and IPAG: Institut de Plan\'etologie et d'Astrophysique de Grenoble, 
  Universit\'e Joseph Fourier, Grenoble 1/CNRS-INSU, UMR 5274, 38041
  Grenoble, France
\and European Space Agency, ESTEC, Keplerlaan 1, 2201 AZ Noordwijk, The Netherlands
\and Department of Physics, Princeton University, Princeton, New Jersey, USA
\and CNRS, IRAP, 9 Av. colonel Roche, BP 44346, 31028 Toulouse Cedex 4, France}  


 \abstract
 {The \Planck\ High Frequency Instrument (HFI)   surveyed 
the sky continuously  from August 2009 to January  
2012. Its noise and sensitivity performance were excellent (from 11 to 40 aW Hz$^{-1}$), 
but the rate of cosmic-ray impacts on the HFI detectors was 
unexpectedly higher than in other instruments. Furthermore, collisions of cosmic rays with the 
focal plane produced transient signals in the data (glitches) with a wide range of characteristics
and a rate of about one glitch per second. A study of cosmic-ray impacts on 
the HFI detector modules has been undertaken to categorize
and characterize the glitches, to correct the HFI time-ordered data, and understand the residual 
effects on \Planck\ maps and data products.
This paper evaluates the physical origins of glitches observed by the HFI detectors.
To better understand the glitches observed by HFI in flight, several 
ground-based experiments were conducted with flight-spare HFI bolometer modules.  
The experiments were conducted between 2010 and 2013 with HFI test 
bolometers in different configurations using varying particles and impact energies.  
The bolometer modules were exposed to 23\ MeV protons from the Orsay IPN Tandem accelerator, and to $^{241}$Am and $^{244}$Cm  $\alpha$-particle and $^{55}$Fe radioactive 
X-ray sources.  The calibration data from the HFI ground-based 
preflight tests were used to further characterize the glitches and compare glitch 
rates with statistical expectations under laboratory conditions.
Test results provide strong evidence that the dominant family of glitches observed in 
flight are due to cosmic-ray absorption by the silicon die substrate on which the HFI detectors reside.
Glitch energy is propagated to the thermistor by ballistic phonons, while thermal 
diffusion also contributes. The average ratio between the energy absorbed, per glitch, in the silicon die and that absorbed in the bolometer is equal to 650. We discuss the implications of these results for future satellite 
missions, especially those in the far-infrared to submillimeter and millimeter regions 
of the electromagnetic spectrum.}
\titlerunning{Impact of particles on the \Planck\ HFI detectors}

\maketitle

\section{Introduction} \label{sec1}

The \Planck\ mission \citep{Planck2013gen}\footnote{\Planck\ (\url{http://www.esa.int/Planck}) is a project of the European Space Agency (ESA) with instruments provided by two scientific consortia funded by ESA member states (in particular the lead countries France and Italy), with contributions from NASA(USA) and telescope reflectors provided by a collaboration between ESA and a scientific consortium led and funded by Denmark.} has observed the sky between August
2009 and August 2013 in the frequency range from 30\,GHz to 1\,THz from its orbital vantage 
point about the second Sun-Earth Lagrange point (L2). The \Planck\ satellite comprises a
telescope, a service module, and two instruments: the High Frequency
Instrument (HFI) and the Low Frequency Instrument (LFI). The HFI operates
with 52 high-impedance bolometers cooled to 100\,mK in a range of frequencies 
between 100\,GHz and 1\,THz and shows an unprecedented photon sensitivity within 
its frequency bands \citep{Planck2011perf}. At the same time, however, the 
HFI detectors exhibit a strong coupling with cosmic-ray radiation that 
produces transient glitches in the raw time-ordered information (TOI). 
The data affected by large-amplitude glitches are rendered unusable 
for science and must be masked in subsequent data processing. 
The redundancy of \Planck's scanning strategy ensures complete sky coverage even 
after large-amplitude glitch masking \citep{Planck2013glitch,Planck2011dpc}. 
Smaller-amplitude glitches are flagged and the data are corrected by 
the subtraction of a parametric fit to an HFI glitch template.  Even smaller glitches could remain hidden 
within the TOI noise, yielding an additional non-Gaussian systematic effect. 
Therefore, a proper understanding and careful data processing of the glitches are essential to meet the precision 
cosmology goals of the \Planck\ mission.  Uncorrected glitches can 
influence the estimation of the cosmic microwave background (CMB) angular power
spectrum; in particular, there is a potential confusion between glitch residuals and the 
non-Gaussian features caused by topological defects 
or inflation processes \citep{Masi2010}.

The impact of cosmic-rays on the time-ordered data has also been observed in previous far-infrared space missions that used bolometers. For example, glitches in the COBE-FIRAS data were identified to be caused by cosmic-particle hits on the detectors, because they were not correlated to the position of the mirrors (\citep{Boggess,Fixsen}). The number of glitches observed for this experiment was sufficiently small and their removal was not a major problem. Glitches have also been identified in the Herschel space observatory, which was injected in a Lissajous orbit around the L2 point together with the \Planck\ satellite. It contains the SPIRE photometer, which uses three bolometer arrays of hexagonally close-packed feedhorn-coupled NTD-detectors operating at 300 mK (43 at 500 $\mu$m, 88 at 350 $\mu$m, and 139 at 250 $\mu$m) \citep{Griffin}.  Glitches caused by cosmic rays have been observed in SPIRE data. Two types of glitches are seen in the SPIRE detector timelines: large events and smaller co-occurring glitches both associated to the impact of cosmic rays on the arrays.

This work, together with a companion article \citep{Planck2013glitch}, provides an improved understanding of
the physical origin of the glitches observed in the HFI instrument
in-flight data. This paper discusses several 
ground-based experiments using the collision of protons and $\alpha$-particles
with spare HFI bolometers.  These experiments provide better control of the
incident particle characteristics (e.g., particle type, energy, impact location) and 
environmental conditions than the HFI in-flight data.

This article is structured as follows: we first give a short review of
the cosmic-ray flux at the second Sun-Earth Lagrange point (L2). In Sect.~\ref{sec3} we describe the geometrical and thermal
characteristics of the elements that constitute an HFI bolometer module. In
Sect.~\ref{sec4} we summarize the results obtained from the analysis
of the glitches in the HFI in-flight data. In
Sect.~\ref{sec_ground} we show the results of the ground
characterization of the HFI spare bolometers and describe several radiation tests
with protons and $\alpha$-particles. In conclusion, we discuss the
physical interpretation of the glitches seen by HFI in flight and the
lessons learnt for future cosmological space missions.

\section{Cosmic rays at the second Lagrange point}\label{sec2}

Cosmic rays (CRs) at L2 are composed of about 89\,\% protons, 10\,\%
$\alpha$-particles, and 1\,\% nuclei of heavier elements; less than
1\,\% are electrons \citep[see, e.g.,][]{Mewaldt2010,Leske2011}.  The
most important contribution (in terms of both energy and quantity) to
the CRs at L2 comes from within our own Galaxy.  The incoming charged
particles are modulated by the solar wind, which decelerates and
partially shields the inner solar system from lower-energy galactic
CRs. There is a significant anticorrelation between solar activity and
the flux of CRs with energies below 10\,GeV.

In addition to deflecting galactic CRs, the Sun is itself a source of cosmic-ray nuclei and electrons that are accelerated by shock waves that travel
through the corona, and by magnetic energy released in solar
flares. The typical energy reached in solar particles is in the keV range. On the other hand, during a solar flare, the highest energy reached is typically 10 to 100\,MeV, occasionally reaching 1\,GeV (roughly once a
year) to 10\,GeV (roughly once a decade). 
During the \Planck\ mission, we
observed 13 solar flares; each solar flare lasted a few hours and
produced a change in temperature of the whole bolometer plate
that affected the working point of the bolometers. The data related to
these events are flagged and excluded from standard HFI data processing, therefore  
we did not consider this source of CRs here.

A third population of CRs, referred to as \emph{anomalous cosmic rays} (ACR), 
consists primarily of photoionized interstellar hydrogen nuclei that are 
accelerated at the termination shock in the heliosphere \citep{Stone2012}.  
Although heavier elements are present in the ACR population, protons with energies from 
hundreds of keV to 100\,MeV represent a marginally important component of the 
cosmic-ray flux that affects the \Planck\ HFI. 

The flux of CRs is monitored onboard the \Planck\ satellite by the
Standard Radiation Environment Monitor (SREM)\,\footnote{SREM consists of three detectors (diodes D1, D2, D3) 
in two detector head configurations \citep{SREM2003}. 
A total of 15 discriminator levels are available to bin the energy of the detected events.} 
mounted on the exterior of the spacecraft.
Solar flares provided a useful test to correlate the signal measured on the outside 
of the spacecraft with the SREM with signals due to particle impact on HFI. We found that only CRs with energies above 39\,MeV can penetrate 
the focal plane unit (FPU) box \citep{Planck2013glitch}.

\begin{figure}
\begin{center}
\includegraphics[width=\hsize]{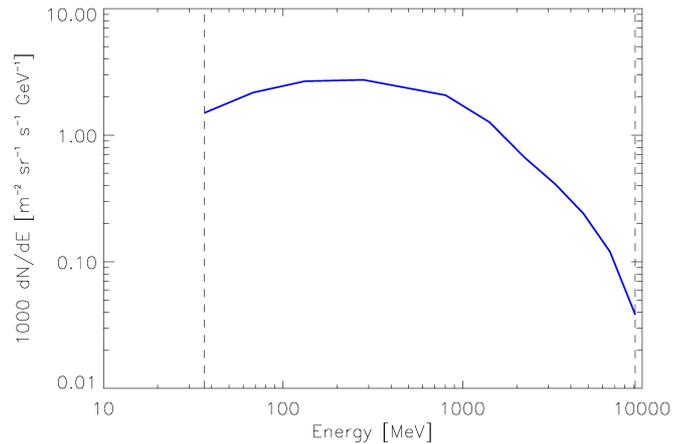}
\end{center}
  \caption{Energy distribution of CR protons at L2. The vertical dashed lines represent the range of
    energies of interest for HFI (from 35\,MeV to 10\,GeV).}
\label{fig1}
   \end{figure}

We conclude that the flux of particles that can cause glitches
in the HFI data is composed of protons in a range of energies between
39\,MeV and 10\,GeV. The energy distribution of protons is presented in
Fig.~\ref{fig1} \citep{Adriani2011,Christian2011,Picozza2011}. 
Below 39\,MeV, protons do not have enough energy to penetrate the FPU
box to reach the 100\,mK stage. The total flux of CRs
peaks around 200\,MeV, giving a total proton flux of 
3000\,--\,4000\,particles\,m$^{-2}$\,sr$^{-1}$\,s$^{-1}$\,GeV$^{-1}$.

\begin{figure}
\begin{center}
\includegraphics[width=8cm]{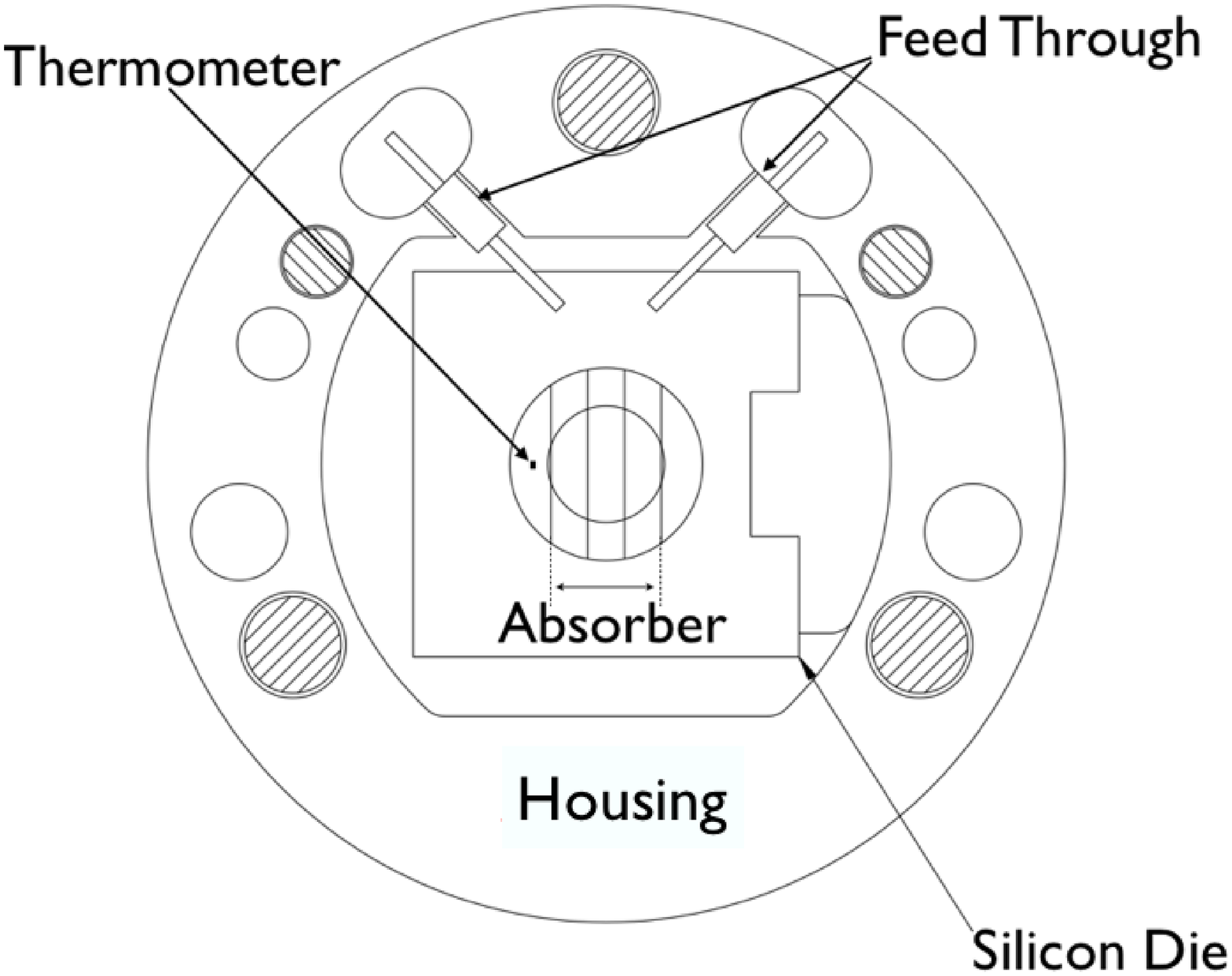}\\
\includegraphics[width=8cm]{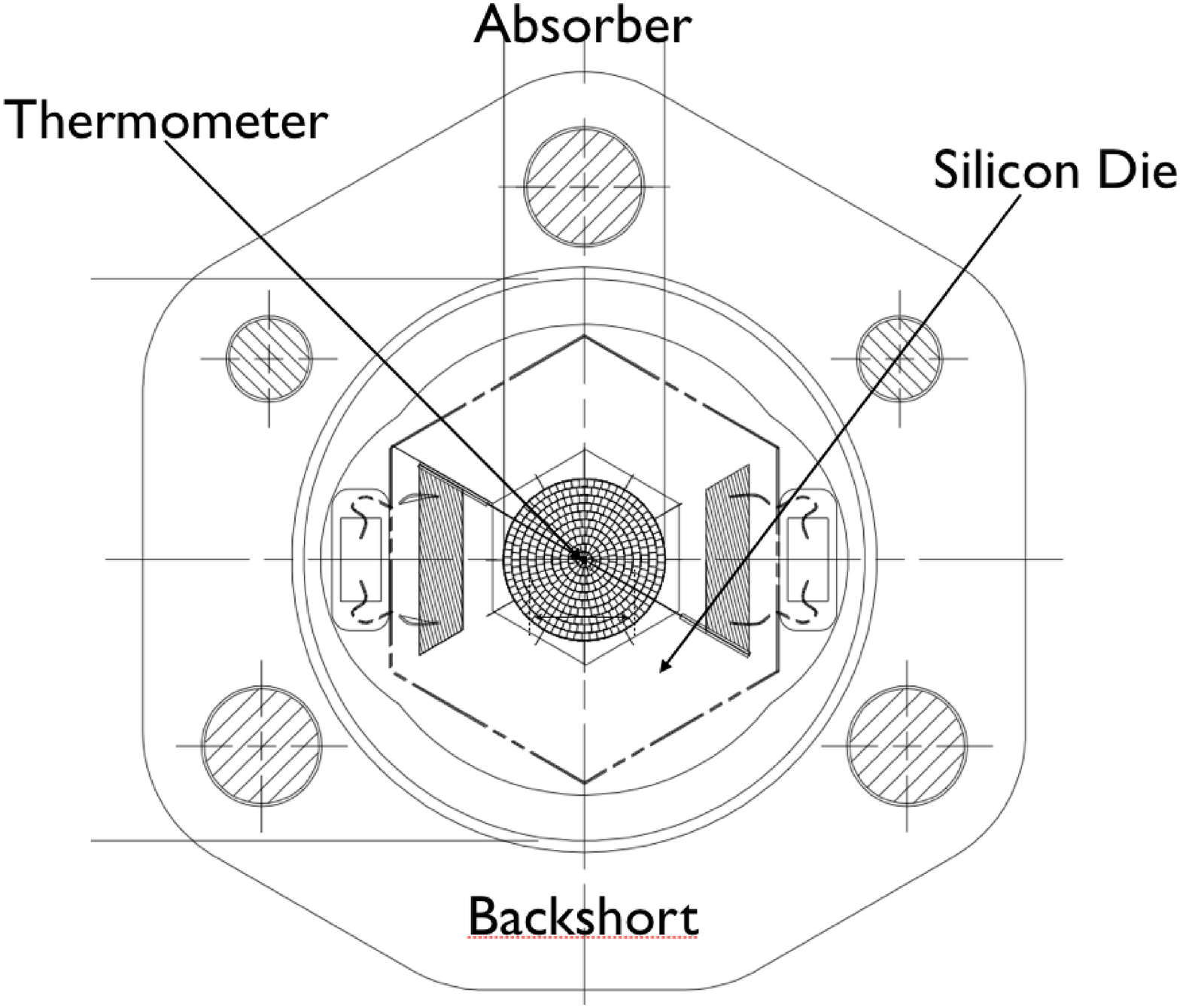}
\end{center}
  \caption{\textit{Top}: front perspective mechanical drawing of a polarization-sensitive bolometer (PSB). The illustration shows the forward/upper portion, i.e., PSBa, of the PSBa/b pair. The orientation of the PSBb absorber is orthogonal to the PSBa.  \textit{Bottom}: mechanical drawing of a spider-web bolometer (SWB). In both figures, the printed wiring board (PWB) and cover are not shown. The size of a HFI bolometer module (including the thermometer, the absorber, the silicon die, the housing and the backshort) is between 1.5 and 1.9\,cm depending on the working frequency.}
\label{fig2}
\end{figure}

\section{HFI bolometers}\label{sec3}

In this section we describe the different elements that constitute an
HFI bolometer module. In general, each element of a bolometer module
can interact with CRs and, through the thermal link with the 
thermistor, it can generate a glitch in the data. One of the goals of
this article is to isolate the elements of an HFI bolometer that can
have an impact on the HFI data.

The HFI high-impedance bolometers \citep{Holmes2008} are
manufactured using silicon nitride supported metal-mesh absorbers and neutron
transmutation doped (NTD) germanium thermistors with a resistance strongly
dependent on the temperature.  The use of an absorber with a spider-web shape strongly reduces the cross-sectional area ({by about 95\,\%}) for CR and particle collisions and it was a design consideration for detectors intended for on-orbit and balloon-borne applications.  At the same time, 
the effective in-band cross-sectional surface area is similar to that of a solid absorber; 
thus, the spider-web absorber technology reduces the CR and out-of band noise produced in the TOI data.  
For reviews of the theory of high-impedance
bolometers see \cite{Jones1953}, \cite{Mather1982}, \cite{Mauskopf1997PhDT}, \cite{Sudiwala2000}, and \cite{Vaillancourt2005}.

The size and grid spacing of the mesh is adapted to operate in six separated frequency
bands between 100 and 857\,GHz. In flight, four spider-web bolometers (SWBs)
in each of the five bands from 143 to 857\,GHz measure the total intensity of incident radiation.
 In addition, eight polarization-sensitive bolometers (PSBs) in each of four bands
(100, 143, 217, and 353 GHz) couple only to a single linear
polarization and thus measure a linear combination of Stokes
parameters $I$, $Q$, and $U$.
PSBs are arranged in pairs sensitive to orthogonal
polarizations. Bolometer performance strongly depends on the read-out
electronics design \citep{CatalanoACDC2010}. The electronics chain implemented
in the HFI is an evolution of the square-wave-biased AC readout
\citep{rieke,Gaertner1997}, known to improve the low-frequency stability of bolometers.

The elements of an HFI bolometer module are the following:
\begin{itemize}
\item \emph{Thermometer}: the bolometer temperature is measured using
  an NTD germanium semiconducting
  thermistor. The size of the NTD thermometers is the same  in all the SWBs
  and PSBs; it is 30 (thickness) $\times$ 100 $\times$ 300\,$\mu$m.

\item \emph{Absorber}: the millimetre-wave absorber for SWBs and PSBs is a
  free-standing metallized Si$_3$N$_4$ micromesh supported by Si$_3$N$_4$ beams. The
  width of the micromesh varies with the HFI band-frequency;  
  it is between 5 and 10\,$\mu$m. The 1\,$\mu$m thickness of the 
  Si$_3$N$_4$ structures is the same in all the bolometers.   In the PSBs,
  the grid spacing, always $>50$\,$\mu$m, is chosen so that
  it absorbs millimetre-waves with high efficiency but has a much smaller
  physical surface area, which significantly reduces the cross-section to cosmic-ray
  particles and shorter-wavelength photons.

\item \emph{Silicon Die}: this is a silicon wafer with an initial diameter of
  15.2\,cm, diced differently for SWBs and PSBs (see
  Fig.~\ref{fig2}). A Si$_3$N$_4$ film of 1\,$\mu$m was deposited on both 
  sides of the wafer.  Gold with a titanium adhesion layer was
  deposited and patterned on the side of the wafer to define the
  millimetre-wave absorber, the thermal link to the absorber, and the
  thermalization bars and contact pads for wire bonding and the indium
  bump bonds. The bolometer die is located over the backshort (or the housing for the forward PSBs) and
  glued with epoxy. The silicon die thickness is  350\,$\mu$m in all 
  the bolometers and the surface area is between 0.4 and 0.8\,cm$^{2}$,
  depending on the working frequency.

\item \emph{Backshort}: made of beryllium-copper, it supports the silicon die
  and permits an optimal absorption of photons by the grid as a result of 
  the $\lambda/4$ relief machined into it. It is present in SWB bolometers and 
  in the rear PSBs (PSBb).

\item \emph{Housing}: made of beryllium-copper, it supports the silicon die of the PSBa bolometers. 
	
\item \emph{Printed wiring board (PWB)}: the PWB is mounted in the
  bolometer module on the opposite side of the backshort from the
  bolometer and contains surface mount inductors and a ground plane
  that, with the capacitors and bolometer, forms a filter to attenuate
  radio-frequency signals before it is rectified at the bolometer.

\item \emph{Feed-through}: this element is only present in the
  forward bolometers of a PSB module (PSBa). It 
  electrically links the gold pad of the forward PSB to the module PWB with a very
  low level of electrical cross-talk with the backward PSB (PSBb). This is achieved by using an
  aluminium cylinder glued with epoxy in the backshort of the backward PSB. Inside the
  aluminium the wire is suspended with epoxy.

\item \emph{Cover}:  the cover is made of beryllium-copper. It is installed onto
  the backshort to protect the PWB and to support a connector.

\end{itemize}

\begin{table*} 
  \caption{Summary of the ground-based bolometer tests}
  \label{tab2}
  \centering
  \begin{tabular}{lcccc}
  \hline\hline
  & Tandem acc. Tests&  $\alpha$ Test 1 & $\alpha$ Test 2 & HFI Ground Cal. \\
  \hline
  Location  &IPN, Paris        & N\'eel Institute, Grenoble& IAS, Paris& CSL, Li\`ege \\
  Period     &Dec 2010 			      & Nov 2011--Apr 2013& Jun--Nov 2011& Jun--Jul 2008 \\
  Source	 & Tandem accelerator           &$^{55}$Fe isotope source (X-rays 5.9\,keV) & $^{241}$Am source (3\,Bq)&Secondary CR emission \\
                 &$p^+ $ at 23\,MeV            & $^{244}$Cm source ($\alpha$ particle at 5.9\,MeV)& $\alpha$ particles at 5.4\,MeV &  \\
  Cryostat	 &N\'eel 100\,mK dilution 					& N\'eel 100\,mK dilution& IAS 100\,mK dilution & HFI Cryostat\\  
  Detectors	 &3 SWB 	& 1 SWB, 1 PSB& 1 SWB, 1 PSB &  52 HFI flight bol. \\
  Read-Out El.& AC biased (same as HFI) 			&AC biased (same as HFI) 	& DC (dig. at 5\,kHz)& HFI AC biased \\
  \hline
  \end{tabular}
\end{table*}

\begin{table*} 
\caption{Energy absorbed [keV] in the different elements of a
  bolometer module for  normally incident particles}
\label{tab1}
\centering
\begin{tabular}{lcccccc}
\hline\hline
  Particle &NTD & Grid & Silicon die & Feed-through\\
  \hline
  30\,MeV--10\,GeV $p^+$ (CR@L2) &	14.7--75.8 & 0.5--4.7	& 118--640 & 200--1000 \\
  23\,MeV $p^+$ (Tandem acc.)      & 137	& 6.5 & 1600& Not reached&  \\       
  5.9\,MeV $\alpha$ ($^{244}$Cm)  &  Stopped &890 & Stopped& Not reached \\
  5.4\,MeV $\alpha$ ($^{241}$Am)  & Stopped & 950& Stopped & Not reached  \\  
 \hline
\end{tabular}
\end{table*}

\section{HFI in-flight glitches}\label{sec4}

The structure and evolution of the in-flight glitches
 have been discussed  in \citet{Planck2011dpc} and  the companion article
\citet{Planck2013glitch}. In this section we summarize
the principal observed characteristics of the in-flight HFI glitches.

We describe the glitches observed by HFI in terms of three characteristics: 1) rate of events and its
evolution during the mission; 2) characteristic event profile (glitch template);
3) the absorbed energy distribution.

\begin{figure}
\begin{center}
\includegraphics[width=\hsize]{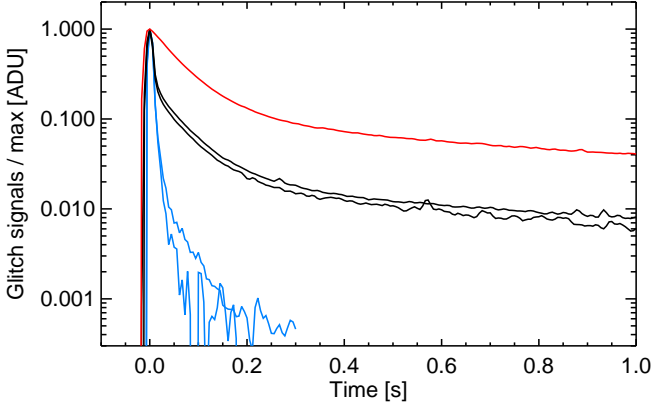} 
\end{center}
  \caption{HFI flight results: average short glitch template (blue),
    long glitch template (black), and slow glitch template (red) for one PSB in-flight
    bolometer \citep{Planck2013glitch}.}
\label{fig3}
   \end{figure}  

\begin{figure}
\begin{center}
\includegraphics[width=\hsize]{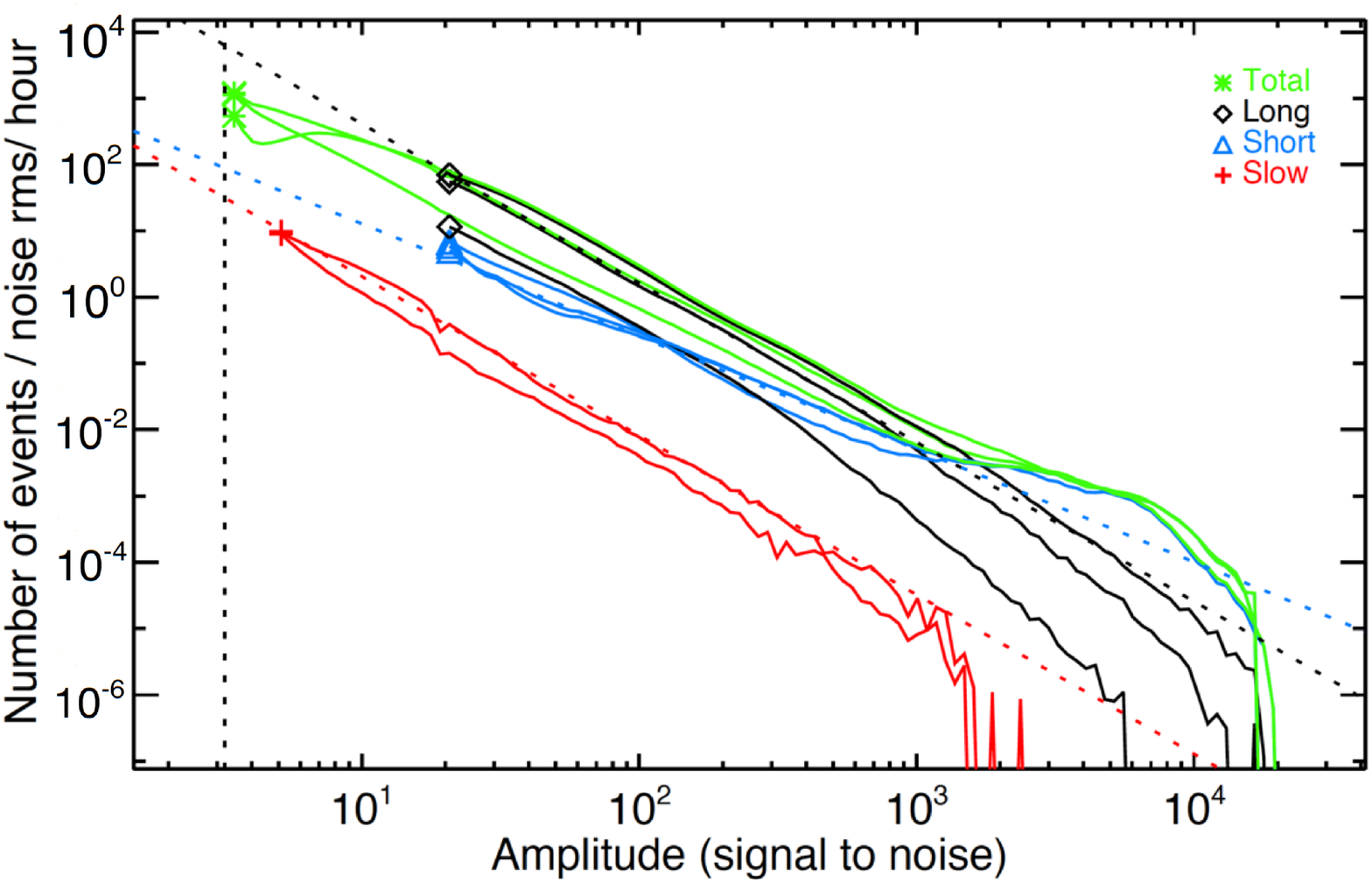}
\end{center}
  \caption{ HFI flight results: glitch energy distribution for the three families of glitches
with respect to the peak amplitude, in signal-to-noise units, for three different PSB and SWB in-flight bolometers \citep{Planck2013glitch}. 
  The blue line is for the short glitch population, black is for long, red is for slow, and green is for total. Power laws are shown for comparison as dashed lines.}
\label{fig4}
\end{figure}  

 \begin{figure}
 \begin{center}\label{tandem}
\includegraphics[width=8cm]{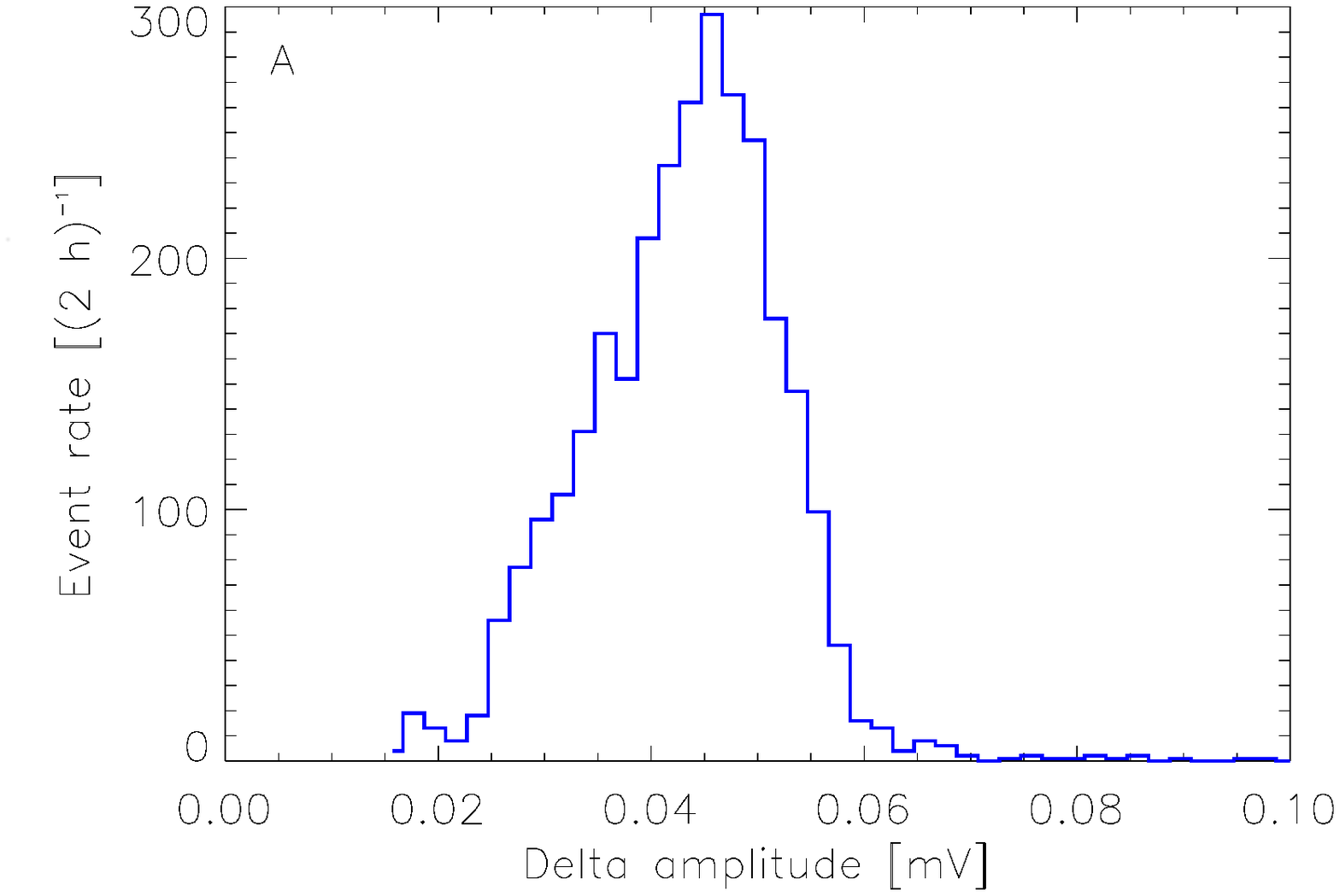}
\includegraphics[width=8cm]{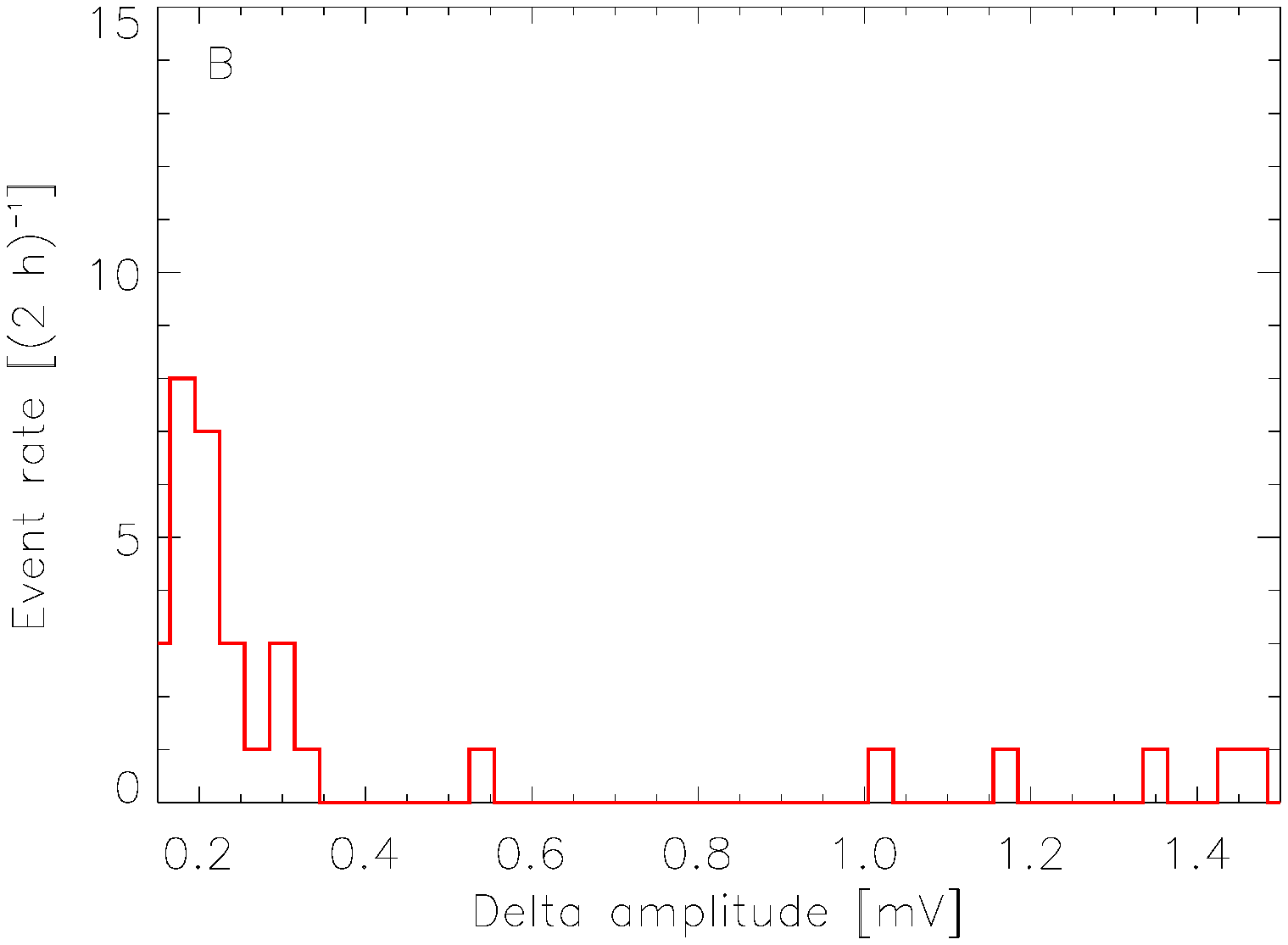}
\includegraphics[width=8cm]{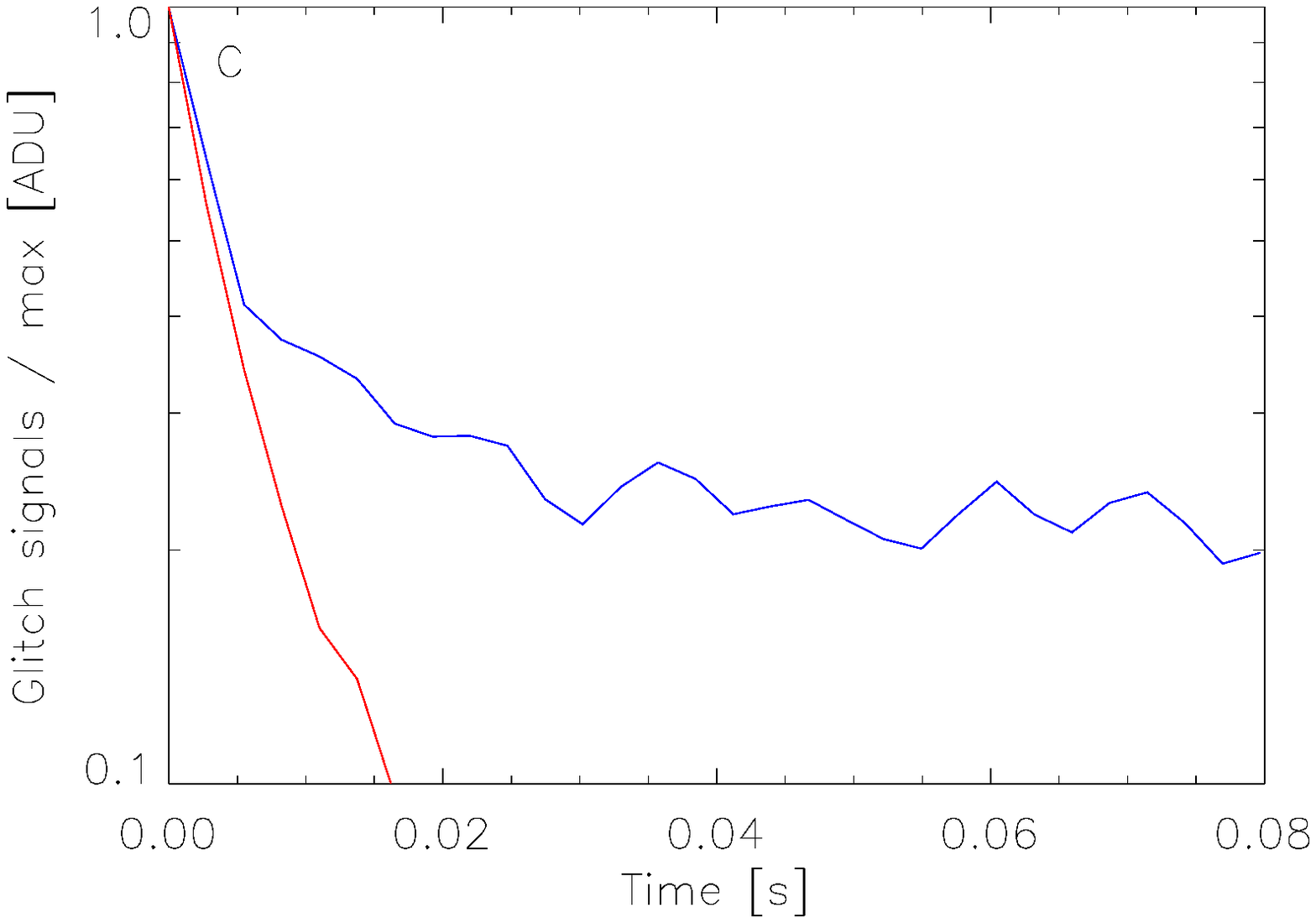}
\end{center}
  \caption{Tandem accelerator test. (A) Histograms of the glitch amplitudes
   in 2\,h of integration time for low
    amplitudes (0\,V\,--\,0.1\,mV). (B) Histograms of the glitch amplitudes
    in 2\,h of integration for
    high amplitudes (greater than 0.1\,mV). (C) Glitch template
    built by stacking all the events of the first family centered at
    45\,$\mu$V (blue) and the family centered at 0.15\,mV (red). Both templates show the same
fast time-constant $\tau$ (4\,ms), and the blue template shows a slower time-constant  (about 20\,ms) with an amplitude relative to the peak of few percent.}
\label{fig5}
   \end{figure}

\begin{itemize}
\item \emph{Rate}: The evolution of the glitch rate is strongly
  correlated with the SREM as
  discussed in Sect.~\ref{sec2}. This is evidence that glitches are produced
  by CRs and not by local phenomena such as vibrations or read-out
  electronics noise. The measured in-flight glitch rate was about 20 times
  higher than expected from the coupling between the absorber and the
  CRs at L2. A phenomenon other than the direct interaction of the CRs
  with the absorber and NTD has to be invoked to explain this
  \citep[see][]{Planck2013glitch}. 
  Two hypotheses have been put forward: (1)  the
  NTD thermometer is sensitive to a change in temperature of other
  (larger) elements; (2)  a large portion of the
  glitches come from indirect interaction between CRs and bolometers,
  for example, protons that interact very close to the surfaces of
  materials surrounding the NTD+Grid (in particular copper) produce
  electron showers able to propagate to the absorber+NTD.

\item \emph{Shape}: From data reduction  \citep{Planck2013glitch}, we distinguished
  three families of gltches with different shapes (see
  Fig~\ref{fig3}): short, long, and slow 
  glitches. We can fit each family with a template that is the sum of
  first-order low-pass filters:
\begin{equation}
{\rm template} = \sum_{i=1}^N  A_i \cdot \exp(- t/\tau_i).
\end{equation} 
Short glitches present a fast decay (between 4 and 10\,ms, depending on
the bolometer) and a tail with an amplitude of a few percent of the
maximum.  
Long glitches present the same fast decay as the short ones, but have a
long (about 10\,\%) relative amplitude tail that shows intermediate to 
long time-constants of tens of milliseconds and about
1\,s. 
Slow glitches show the same slow time-constants as the long and short glitches, but not the fast decay.
The residual glitch signal after template subtraction is significantly reduced.
Indeed, the remaining contamination from glitches is below the instrumental noise level, even for channels with the highest glitch rates. 

\item \emph{Spectrum}: The glitch energy distribution was
  estimated for short, long, and slow glitches. The distributions can
  be fitted by power-laws with different indices, and different bolometers show very similar
  results. In Fig.~\ref{fig4}, we present the
  averaged energy spectra of the  in-flight PSBs at
  217\,GHz collected during the full mission. The total number of
  glitches 
  is dominated by long glitches at low absorbed energies. Short
  glitches dominate for high absorbed energies. Slow glitches are
  present only on the PSBa bolometers. Long and slow glitches show
  the same power law index. The short-glitch distribution shows a double
  structure with an elevated bump in the high-energy range.

\end{itemize}
 
\section{Ground measurements}\label{sec_ground}

For an improved understanding of the origins of the glitches seen in
HFI flight data, we performed several ground-based tests on spare HFI
bolometers between December 2010 and May 2013. The aim
of the tests was to obtain a more complete view of the physical origin of the
different families of glitches. These tests were performed in
different configurations using 23\,MeV protons from an ion
accelerator and two radioactive $\alpha$ particle sources ($^{241}$Am and
$^{244}$Cm). In addition, we used the HFI ground-calibration data to
check the impact of the cosmic rays at sea level with HFI
bolometers. The sensitivity reached for these tests was about 10\,nV\,Hz$^{1/2}$, very close to the flight performance.
A list of all the tests that were performed is presented in Table~\ref{tab2}.

We started by placing spare HFI SWBs in front of an ion accelerator
with a proton beam able to cover the full surface of the
bolometer module. The results of this test showed that we are able to
reproduce short and long glitches observed in flight without 
any delta-electron\footnote{Electrons ejected from matter by ionizing radiation.} 
production process. Because of the limited
integration time during the accelerator test, we performed other tests with $\alpha$ sources 
to investigate the production of long glitches more deeply by using a
collimated  beam able to interact with small parts of the
bolometer module on each test. The results of these tests 
showed that the silicon die is the origin of long
glitches. Two hypotheses were put forward to explain the heat propagation from the silicon die to the NTD thermometer:  ballistic phonon propagation and ionization. To 
distinguish between these two effects we performed a test by breaking 
the Si\,$_3$N\,$_4$ legs that support the absorber (65\,\% of the legs were severed) to reduce the ballistic heat transport by a measurable factor. 

In the next subsections we give the principal results of each test;
this motivates the physical interpretation of glitches in HFI data
discussed in Sect.\ref{sec:implications}.

\subsection{Particles on HFI detectors}

In the range of energies of interest, the main interaction between
particles and materials is ionization. The value $\Delta E / \Delta x$
versus particle energy is given by the total stopping power curves
for silicon, germanium, and copper, and corresponds to the sum
of the electronic stopping (inelastic collisions between bound
electrons in the solid medium and the ion moving through it) and the
nuclear stopping (elastic collisions between the ion and atoms in the
sample, almost negligible in the range of energies of interest).
In Table~\ref{tab1}, we present the energies absorbed in the relevant
elements of a bolometer module from $\alpha$ particles and protons at normal incidence.

\subsection{Tandem accelerator test}

This test was performed in December 2010 at the Institut de Physique
Nucleaire (IPN -- University of Paris-sud) using the Tandem
accelerator\footnote{\url{http://ipnweb.in2p3.fr/tandem-alto/}}: this accelerates ions by applying an electrostatic field with a feature of supplying high voltage in the
  middle of the accelerating tube. It can accelerate ions in two steps:
  first, is accelerates negative ions and second, it accelerates
  positive ions to higher velocities after transforming negative ions into positive ions in the
  high-voltage terminal ($E=(q+1)V$ [MeV]). 
The Tandem accelerator
can produce 23\,MeV protons with a beam of about 2.5--3\,cm always
orthogonal to the detectors. This means that the beam
covers the full surface of the bolometer module.

The aim of this test was to obtain a general view of the interaction
between protons and the different elements of an HFI bolometer
module. We used three spare HFI SWBs. In front of one, at a distance
of about 5\,mm, we set a 10\,$\mu$m gold shield to test for possible
delta-electron production. The second was placed in front of the
Tandem beam without any shield or optics (filters, horns, etc.). The
last was used as a reference bolometer.

In Fig.~\ref{fig5}, we present the results of the test for the unshielded bolometer. 
The histogram of the highest glitch amplitude is shown.  Two families of events
can be isolated. The first family peaks at 45\,$\mu$V and contains
98.5\,\% of the glitches. The template built by
stacking all the glitches of this family (blue curve in Fig.~\ref{fig5}C)
 can be represented by a fit of a two-time-constant model (a
faster $\tau$ at 4\,ms and a slower $\tau$ at about 20\,ms). This family is similar to
the long-glitch family seen within HFI in-flight data. The second family of glitches peaks at
about 0.15\,mV and contains about 1.5\,\% of the glitches. 
These glitches agree reasonably well with a template fitting with a single time-constant
model with the same 4\,ms time-constant as the first glitch family (red
curve in Fig.~\ref{fig5} - C). This family of glitches is similar to the short-glitch family seen in HFI 
in-flight data. The limited
integration time of Tandem experiment did not provide a sufficiently high signal-to-noise ratio to allow a confident fitting of the longer time-constants, as is 
observed in HFI in-flight data.
All the glitches have rising times faster than the sampling of the
AC read-out electronics, therefore in most cases the peaks correspond to the
first sampled glitch data point. 

The bolometer with a gold shield shows the same results as the one
without the shield. It seems that we are not able to observe any delta
electrons under these conditions.

We can summarize the results of the Tandem accelerator test as follows:

\begin{itemize}

\item We are able to isolate long and short glitches without
  introducing any delta-electron production process.

\item The two families of glitches show the same rate ratio and the
  same time-constants as observed in HFI in-flight data.

\item Slow glitches were not observed because we did not use PSB
  bolometers during this test.

\end{itemize}

\begin{table} 
\caption{Calibration factors for the 217\,GHz SWB, measured under various operating conditions}
\label{tab3} 
\centering
\begin{tabular}{lcccc}
\hline  \hline
&Run 1 & Run 2 & Run 3 & Run 4 \\
\hline
$T_{\rm plate}$	[mK]  &80 & 80 & 80 & 80 \\  
$F_{\rm mod}$ [Hz] &200 & 90 & 182 & 90 \\
$I_{\rm bias}$ [nA] & 0.2 &  0.4 &  0.4 &  0.2 \\
Cal-fact [$\mu$V/keV] & 67.5 & 20.0 & 28.0 & 42.2 \\
\hline
\end{tabular}
\end{table}

\begin{figure}
\begin{center}
\includegraphics[width=\hsize]{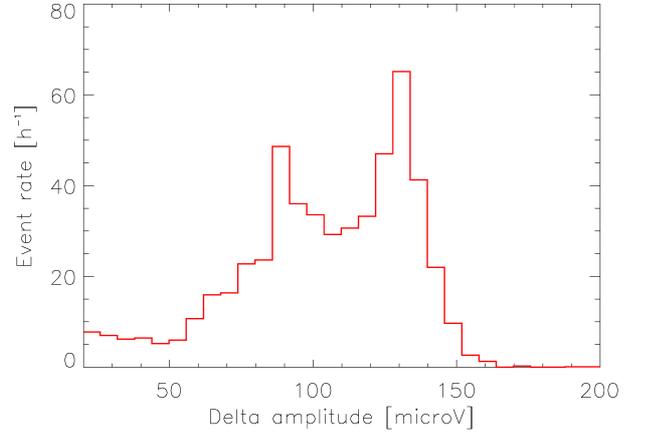}
\end{center}
  \caption{$^{55}$Fe X-ray calibration run. Event rate versus signal. The distribution is asymmetric
and two overlapping peaks are evident.}
\label{fig:cal_run}
\end{figure}  

\subsection{Alpha-particle tests}

These tests were performed between June 2011 and April 2013 
using the same 100\,mK dilution cryostat and read-out
electronics as were used in the Tandem test. The aim of these tests
was to characterize in detail the long glitches observed in HFI in-flight data. The Tandem accelerator tests have demonstrated that it is possible 
to have long glitches without involving any delta-electron creation processes.
Several tests were conducted, as described in the following subsections.

\subsubsection{Calibration run}

We used a $^{55}$Fe source that produced X-rays at 5.9\,keV. A typical
spectrum (event rate versus signal) is presented in
Fig.~\ref{fig:cal_run}. The distribution is asymmetric
and two overlapping peaks are evident. This distribution can be interpreted as follows:
the probability that a 5.9\,keV X-ray
is absorbed by 1\,$\mu$m of silicon nitride is about 3.5\,\%. If
absorbed in the grid, the 5.9\,keV X-ray will ionize the atoms of the
silicon nitride. Some of the photoelectrons produced will be absorbed
in the silicon nitride itself, which means that this energy will be fully
transformed into heat. Nevertheless, as the grid is 1\,$\mu$m thick,
we expect to have a significant rate of photoelectron escape that will
produce an asymmetry in the spectrum at lower signals.  In addition,
at the atomic level, when a core electron is removed and leaves a
vacancy, an electron from a higher energy level may fall into the
vacancy, which results in a release of energy. This energy can be released
in the form of a fluorescence X-ray (for the silicon K$\alpha$/$\beta$
fluorescence X-ray of 1.8\,keV) or the energy can be transferred to
another electron and ejected via the Auger effect.  The Auger
electrons will be absorbed very quickly in the silicon nitride, in
which case all the 5.9\,keV of the $^{55}$Fe X-ray will be fully
transformed into heat and are then measured by the NTD as a change of
temperature. This represents the larger peak in the spectrum at higher
energy. If the absorption process produces fluorescence X-rays of
1.8\,keV (yield of about 5\,\%), they will be able to escape easily
from the silicon nitride, removing 1.8\,keV of the energy to be
transformed into heat.  This represents the second overlapping peak at
lower energy.  In conformity with this interpretation, we chose to
derive the calibration factors in the higher energy peak. Calibration
factors (in $\mu$V/keV) were calculated for several conditions
(see Table~\ref{tab3}).

\subsubsection{Delta electron test}

\begin{figure}
\begin{center}
\includegraphics[width=\hsize]{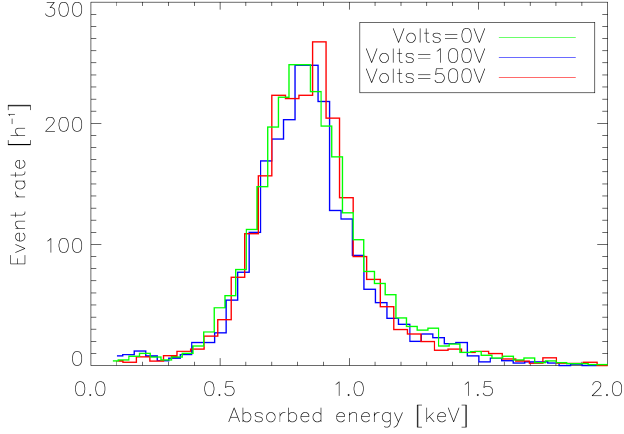}
\end{center}
  \caption{Delta electron test: number of
    events per hour versus signal, with three different applied
    voltages. }
\label{fig7a}
\end{figure}

In this test, the $\alpha$ source was placed in front of the bolometer
with a 2.5\,mm diaphragm at a distance of about 1\,cm. A 10\,$\mu$m
thick copper shield was glued onto the diaphragm. After interacting
with the shield, the $\alpha$ particles retain an energy of about
1.1\,MeV. In this case, the beam spread permits the $\alpha$ particles
to reach both the NTD+absorber and the silicon die. We biased the
source and the copper shield with a variable positive voltage from
0\,V to 500\,V to suppress (or partially suppress) the
contribution from delta electrons pulled out of the shield. In
Fig.~\ref{fig7a}, we present the histograms of the number of counts
versus absorbed energy for the three applied voltages. The three
histograms are centered at about the same value. We conclude that the
ground-based tests did not show any measurable production of
delta-electrons.

\begin{figure}
\begin{center}
\includegraphics[width=\hsize]{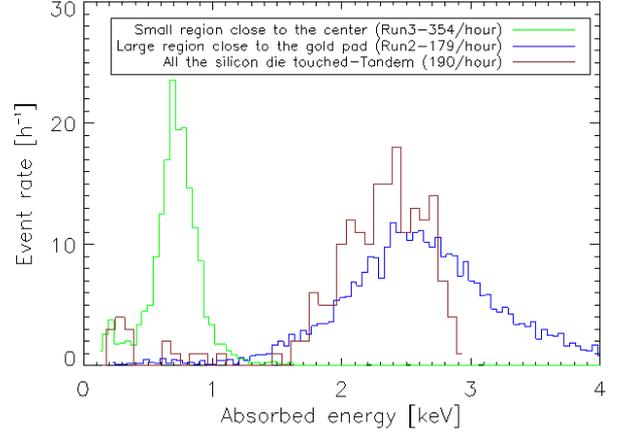}
\end{center}
 \caption{Silicon die tests. Number of events per
    hour versus calibrated absorbed energy in the NTD thermometer for
    two positions of the diaphragm along the silicon die (green
    and blue) and for the Tandem test (brown). The absorbed energies have been
    corrected taking into account the differences of the absorbed
    energy in the silicon die between the tests (1.1\,MeV absorbed in
    the silicon die by $\alpha$ particle in Grenoble tests and 1.6\,MeV
    absorbed in the silicon die from the Tandem accelerator test.)}
\label{fig7b}
\end{figure}  

\begin{figure}
\begin{center}
\includegraphics[width=8.7cm]{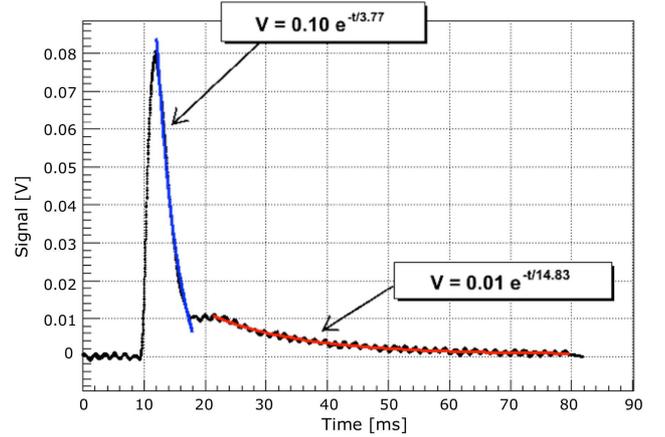}
\end{center}
\caption{Silicon die tests. Glitch template built by stacking 466 events
  caused by impacts in the Si wafer (black curve). The fitted time-constants 
  with exponential models are presented (blue and red curves).}
\label{fig7c}
\end{figure}

\subsubsection{Silicon die tests}

In this test, the $\alpha$ source was placed in a variety of positions in
front of the silicon die. We used a 1.5\,mm diaphragm with 10\,$\mu$m
copper shield at a distance of about 5\,mm from the NTD
thermometer. In this case, the $\alpha$ particles can interact only
with the silicon die.  The 1.1\,MeV $\alpha$ particle is completely
absorbed in the silicon die. We define the attenuation factor
$E_{die}/E_{bolo}$ as the ratio of the energy absorbed per glitch in
the silicon die to the energy absorbed in the NTD thermometer. In this
test we checked the variation of the attenuation factor by moving the
source between two positions along the silicon die.  In
Fig.~\ref{fig7b} we present the comparison between the energy
distribution of the Tandem test (in brown) and the histograms
corresponding to the two positions of the diaphragm. The green line
corresponds to a little region of the silicon die close to the
absorber. The blue line corresponds to larger region close to the gold
pad. The attenuation factor can change from about 500 to 1000 between
the two positions. For Tandem test, where the silicon die
is fully covered by the proton beam, the derived attenuation factor is
650.

In parallel with the tests that used the HFI read-out electronics, we
performed several tests dedicated to characterizing the
interaction of particles with the silicon die using DC readout
electronics operating at ambient temperature without cold
JFETs. This electronics chain has been validated by the IAS team for
testing scintillating bolometers \citep{Coron}. The noise of the
readout was about 11\,nV\,Hz$^{1/2}$, low enough to have a good
signal-to-noise ratio on the glitches.

The aim of these tests was to sample fast
enough to be able to distinguish different physical processes that might
be implicated in the heat propagation from the silicon die to the NTD
thermometer. The results of this test agree with the
Grenoble and Tandem accelerator tests. In Fig.~\ref{fig7c} we present a
typical glitch produced from the hit of an $\alpha$ particle on the
silicon die, along with the time-constants of an exponential model.

\subsubsection{Modified bolometer test}

This test was designed to demonstrate the ballistic phonon propagation
from the silicon die to the grid + NTD thermistor.  To achieve this
goal we broke the grid legs (65\,\% of the leg cross-sectional area
was severed) to reduce the ballistic heat transport by a
measurable factor. The nitride legs were broken using a stainless-steel needle. The break location was made directly at the edge of the Si
frame. The breaks were made by placing the needle next to the leg and
also with the needle next to the Si frame. The needle was then pushed
sideways to break the legs. 

\begin{figure}
\includegraphics[width=\hsize]{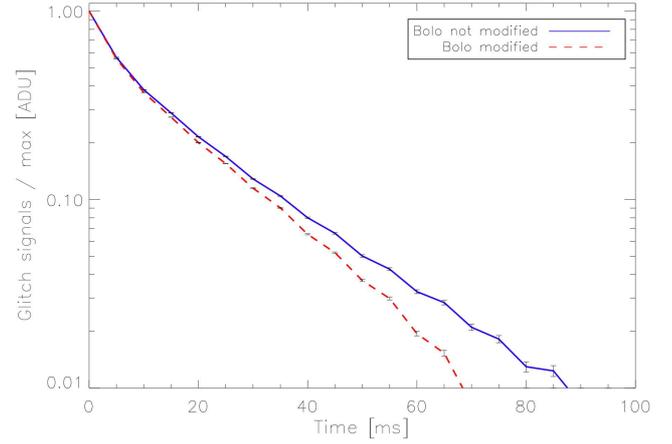}
\caption{Normalized template of the hit on the silicon die. Error bars only show statistical errors.}
\label{fig9}
\end{figure}

We performed an initial reference test with the unmodified
bolometer. Then we performed the test with the modified bolometer,
using the same setup. The spectrum of the count rate versus the
calibrated energy absorbed showed that there was no change in the
measured energy or the counts. We did, however, observe a difference
in glitch shape between the two runs. We built a template by stacking
all the events coming from silicon die hits. We fitted this template
with a two-time-constant model (Fig.~\ref{fig9} and Table~\ref{tab4}).
The amplitude corresponding to the fast time-constant changed by
about 20\,\%. This result is consistent with the hypothesis that part
of the the heat propagates to the NTD thermometer by ballistic
phonons.


\begin{table}
\caption{Effect of removing legs from an SWB: two-time-constant model parameters. The errors include systematic effects between the two runs.}\label{tab4}
\centering
\begin{tabular}{lccc}
\hline  \hline
& Amp$_{\rm fast}$ [\%] &$\tau_{\rm fast} $ [ms] & $\tau_{\rm slow}$  [ms] \\
\hline
Unmodified SWB& $46 \pm 2.1$   &  $4.0 \pm 0.8$  & $20.0 \pm 1.9$ \\  
Modified SWB& $37 \pm 2.6$  & $3.6 \pm 0.9$ & $ 18.9 \pm 1.9$ \\
\hline 
\end{tabular}
  \end{table}

\subsubsection{Coincidence test}

For this test we used a PSB bolometer. The $\alpha$ source was placed in
front of the grids of the PSB using a 1.5\,mm diaphragm with
10\,$\mu$m copper shield at a distance of about 5\,mm from the NTD
thermometer. A PSB bolometer module is surrounded by copper except for
a 1.59\,mm hole in front of the grids and a small aperture on the
silicon die, so in this case, we were completely sure to hit only the
grids.

The coincidence of short glitches on PSBa and PSBb seen in flight
\citep{Planck2013glitch} show that approximately 50\,\% of events are
seen in both PSBa and PSBb without phase shift and without correlation
in amplitude. From modeling the interaction of hundreds of MeV protons with
1\,$\mu m$ silicon nitride with
GIANT-4\footnote{\url{http://wwwasd.web.cern.ch/wwwasd/geant/rd44.html}},
we found that about ten electrons can be ejected from one grid.
This produces a small change in temperature on the other NTD below the
detection threshold. The rate of 50\% correlation can be explained by
geometrical effects such as particles coming from the top or bottom.
Glitch PSBa/b coincidences with a strong amplitude correlation,
on the other hand, which corresponds to a direct impact of protons on the
two grids, were observed to range between 2\,\% and 4\,\%, depending on the
bolometer.

These ground-test results give a level of coincidence of about 3\% for
glitches that are very well correlated in amplitude, when considering
only the glitches detected at or above 5$\sigma$. This agrees
with direct interactions of $\alpha$ particles with the two grids. We
observed only few of glitches without a strong amplitude
correlation, but the integration time of these tests was insufficient
to show the expected 50\,\% level of coincidence due to ejected
electrons. We conclude that ground-based coincidence tests did not
fully explain the coincidence level between PSB bolometers observed in
flight. For a complete analysis of this problem we refer to the
companion paper \citep{Planck2013glitch}.

\begin{figure}[t!]
\begin{center}
\includegraphics[width=9cm]{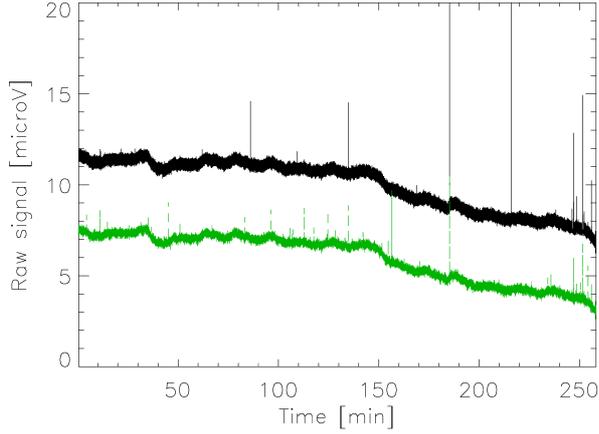}
\end{center}
  \caption{HFI ground calibration. Four hours of time-ordered data (TOD) taken under
    stable environmental conditions. \textit{Black}: PSBa. \textit{Green}: PSBb.}
\label{fig10}
   \end{figure}

\subsection{HFI ground calibration}

During pre-launch \Planck\ system tests, CRs were seen by HFI, but far fewer than at L2. At sea level, muons are the most
numerous charged CR particles, with a mean energy of about 4\,GeV
\citep{ramesh2012}. The integral intensity of vertical muons at sea
level is about 1\,cm$^{-2}$\,min$^{-1}$ for horizontal detectors. The
overall angular distribution of muons at the ground is proportional to
$\cos^2 \delta$, where $\delta$ is the elevation angle.

During the CSL ground-based tests, the focal plane is almost vertical,
therefore we expect to have a few tens of events per hour in the silicon die
and a few per day in the absorber + NTD. The absorbed energy in the
silicon die is about 150\,keV, which means that considering the attenuation factor
measured in the Grenoble tests, muons are expected to produce
measurable glitches during the pre-launch HFI calibration tests. In
addition, these glitches should be seen in
coincidence in the PSB bolometers.

To demonstrate the glitch rate observed during the HFI ground-calibration
testing, Fig.~\ref{fig10} shows a typical PSB that we observed under
stable conditions during the HFI ground-calibration experiments.

We saw about 20 glitches per hour in PSBb (green curve) and fewer in
PSBa (black curve). 
In all the polarized bolometers, the ratio of the
PSBa counts to the PSBb counts is the same; almost all the events seen
in PSBa coincide with an event seen in PSBb.

\section{Discussion}\label{disc}

\subsection{Short glitches}

Short glitches dominate the rate at high energies. This has been
confirmed in the ground-based tests (see, e.g., Fig.~\ref{fig5}). The
thermal response template built from this family agrees well with in-flight HFI bolometer data and ground-based data on the
direct interaction of particles with the absorber and NTD
thermometer. The energy distribution of the short glitches is similar
in all the bolometers and shows a double-peaked structure
\citep{Planck2013glitch}. The bump in the short-glitch distribution,
centered at about 20\,keV, is completely consistent with the
interaction between CRs and the NTD thermometer. At low energies, the
distribution corresponds to the interaction between the absorber and
the CRs.

\begin{figure}
 \begin{center}
 \includegraphics[width=\hsize]{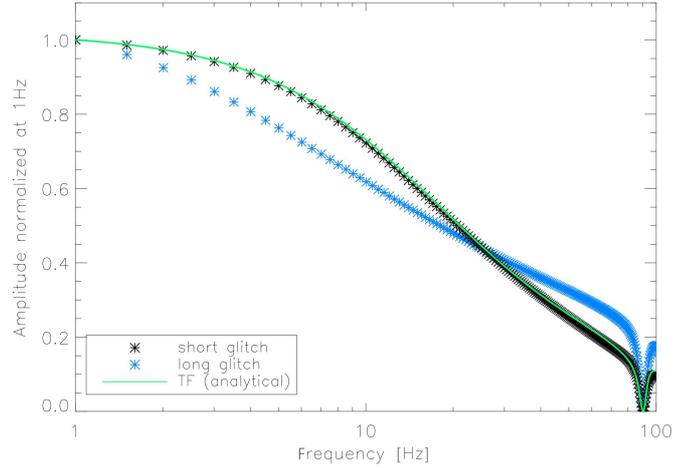} 
\end{center}
  \caption{Comparison of the HFI optical transfer function \citep{Planck2011perf}  (\textit{green line}), short-glitch template (\textit{black stars}), and long-glitch template (\textit{blue stars}).}
\label{fig11}
   \end{figure}

Another indication that short glitches are produced from direct
interaction between CRs and the grid or thermistor can be found in the
comparison between the HFI bolometer electronics time-response
measured on-sky signal \citep{Planck2011perf} and the transfer
function built from the short-glitch template (see Fig.~\ref{fig11}).
The agreement between the transfer function derived from optical measurements
(solid green) and short glitches (black stars) is good, considering
that the template is constructed independently of any model. 
We can conclude that the HFI short glitches are produced by direct interaction
of particles with one of the absorber grids (or directly with the NTD thermistor for a small subset of them).


\subsection{Long glitches}

By contrast, the long-glitch template (blue stars in Fig.~\ref{fig11})
is not consistent with the optical transfer function, indicating a
different physical origin for these glitches.

Two hypotheses have been put forward to explain the many 
long glitches seen in HFI flight data: (1) the NTD thermometer is
sensitive to a change in temperature of other (larger) elements; or
(2) long glitches are caused by indirect interaction between CRs and
bolometers. From the ground-based tests we determined that it is not
possible to generate enough delta-electrons to 
reproduce the energy distribution of HFI in-flight long glitches
(Fig.~\ref{fig4}). On the other hand, we have shown that the NTD
thermometer is sensitive to temperature changes of the silicon
die. All the ground-based tests have confirmed this hypothesis. In
addition, the HFI ground-based calibration data show a rate of events
compatible with the CR flux at sea level over the silicon die
surface. Starting from these considerations, we developed a toy
model of the impact of CRs with the silicon die in orbit at L2 that can explain the long-glitch spectrum seen
by the \Planck\ HFI during flight observations. The
details of this model are presented in \cite{Planck2013glitch}. 
These results agree with the results published by the SPIRE collaboration: small co-occurring glitches that are seen simultaneously on many detectors in a given array in the SPIRE detector timelines, are believed to be due to ionizing hits on the silicon substrate that supports all the detectors in an array \citep{Horeau}.

\begin{figure}
 \begin{center}
 \includegraphics[width=\hsize]{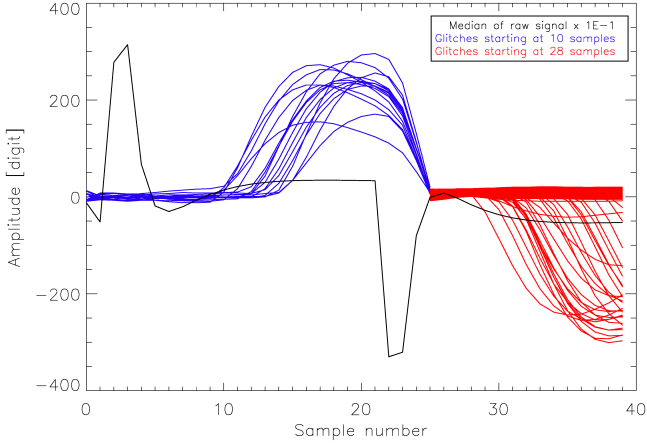} 
\end{center}
\caption{Raw signal analysis of the silicon die tests: unperturbed raw signal (black curve). Difference between a raw signal with a glitch occurred in the center of the half period and the template (blue and red curves).}
 \label{fig12}
 \end{figure}

The physical interpretation of long glitches can be found in the shape
of the long glitches that we have isolated during the ground-based
tests. With the improved temporal resolution made possible by the faster DC-based REU readout electronics
and sampling, we can clearly see (Fig.~\ref{fig7c})
the dual time-constant response that results from heat
transfer between the silicon die and the NTD thermometers.  We
identify two possible hypotheses to explain the heat propagation from
the silicon die to the NTD thermometer:

\paragraph{Ionization and thermal diffusion:}
According to this hypothesis, the faster time-constant is given by ionization
within the silicon die. The ionization signal is obtained by
the collection of the electron-hole pairs created by interaction of
$\alpha$ particles in the silicon crystal via the gold pad. The long
tail can still be explained by the heat diffusion between the silicon
die and the NTD thermometers.
Using the bolometer parameters, we can determine the equivalent bias
step amplitude needed to produce a glitch of about 4\,keV (the typical
energy absorbed in the bolometer from a hit in the silicon die). We
obtain $\Delta I_b = 0.2$\,nA, which is of the same order as the
working bias current. If present, an effect as strong as this should
be evident through inspection of the raw TOI signals. In fact, before
sufficient time has elapsed for heat diffusion within the bolometer, a
collected charge must show a peak in the raw signal with a decay-time
constant determined only by the $RC$ time-constant of the individual
detector read-out circuitry. The sign of this peak should not change
with the changing polarity of the detector bias current. Therefore,
this effect should be easily distinguished from a thermal glitch.  The
results of this analysis are presented in Fig.~\ref {fig12}: in black
we show a template of the typical unperturbed raw signal set during
the test.  The red and the blue curves represent the difference
between a raw signal with a glitch (in the center of the half periods)
and the template. There is no evidence of ionization in the raw
signals; all the glitches can be fully explained by pure thermal
events (i.e., hot and/or cold phonons).

\paragraph{Ballistic phonon propagation and thermal diffusion:}
According to this hypothesis, the faster time-constant, which corresponds to the typical time-constant of the bolometers as 
measured with optical sources in ground- and flight-calibration testing,
can be explained by ballistic phonon heat
conduction:\footnote{Ballistic conduction is the unimpeded flow of
  energy that carries charges over large distances within a material.
}
in our case, ballistic phonons can propagate unhindered from the silicon die and up the grid legs, depending on the geometrical coupling fixed by the size 
and the number of the legs.  The final effect is equivalent to a particle absorbed directly in the grid.

The longer time-constant (tens of milliseconds) represents
the non ballistic thermal heat diffusion between the silicon die and
the NTD thermometers. This implied time-constant is consistent with
the heat capacitance of the silicon die and the thermal conduction of
the silicon die to the bolometer itself. A detailed thermal model of
the bolometer is presented in a companion article (Spencer et al. in preparation)
that describes ground-based tests on flight-spare bolometers conducted
in Cardiff. The tests were performed by placing NTD-Ge thermometers and
100\,k$\Omega$ heaters at various locations on a bolometer module (the
silicon die, the PWB, and the module housing).  Glitch events were
simulated by rapid injection of known quantities of thermal energy
through the additional resistors attached to the bolometer module
while observing the temperature response of the bolometer and its
elements.

The modified bolometer tests of this work have shown that by breaking
8 of the 12 grid-support legs we can reduce the amplitude of
the long-glitch fast time-constant amplitude without an accompanying
decrease in the glitch long-tail contribution.  This implies a reduced
effective ballistic cross-sectional area between the silicon die and
spider-web absorber/NTD. This completely agrees with the
ballistic-conduction hypothesis.

\subsection{Slow glitches}

Slow glitches are the rarest glitch events we see in the HFI in-flight
bolometers. They only affect the PSBa and have a rate of a few per hour in
flight. The energy distribution of the slow glitches shows the same
power-law index as the energy distribution of long glitches (see
Fig.\ref{fig4}). In addition, the slow glitches share the template of
the long glitches, but without the component with the shortest time
constant. The slow glitches were not reproduced during any of the HFI
ground-based tests, both pre-launch with the HFI focal plane unit
(FPU), and post-launch with flight-spare hardware.  In light of this
fact, we are limited to proposing a hypothesis for the origin of the
slow glitches that is consistent with our current results, but lacks
experimental confirmation.  The presence of the feed-through (see
Sect.~\ref{sec3}) connecting the PSBa bolometer to its corresponding
silicon die is the only difference between the PSBa bolometers and the
other types of bolometers in HFI (PSBb and SWB).  The PSBa
feed-through elements have a strong thermal coupling with the silicon
die and gold pad. A proton hitting a PSBa feed-through can therefore 
heat the corresponding silicon die to produce heat diffusion from the
silicon die to the NTD thermometer.  This heating would not have the
corresponding ballistic heat conduction associated with a silicon
die/CR glitch event, and therefore no fast time-constant would be
observed. The differences in the effective surface area of the
feed-throughs with respect to the corresponding silicon dies, of a
factor of about 100, may explain the differences in the rate between
the long and slow glitches.

\section{Implications for future space missions}\label{sec:implications}

\Planck\ HFI has demonstrated the effectiveness of using a grid
absorber instead of a solid/bulk absorber for collecting CMB photons to drastically reduce the number of direct glitches (short
glitches). This characteristic allowed HFI to deliver data at the
signal-to-noise ratio required for the \Planck\ scientific
goals. Nevertheless, we observed more glitches than expected at low energies (long glitches). The ground-based experiments
discussed here demonstrated that the silicon die is able
to propagate the energy absorbed from CRs to the bolometer absorber
and NTD thermometer. In this case, the absorbed energy propagates to
the NTD thermometer via ballistic phonons; this is followed by a
contribution of thermal diffusion from the silicon die to the
absorber and NTD.

The study of the physical processes related to athermal propagation
and ionization in a crystal, and more generally, the effects of
particles in cryogenic bolometers, have several applications, including
new-generation space missions at millimetre and submillimetre
wavelengths \citep[e.g.,][]{Caserta1990}, high-resolution X-ray
spectrometry
\citep[e.g.,][]{Kelley2007,Kilbourne2006,Stahle2004},
and direct detection of dark matter
\citep[e.g.,][]{Sundqvist2008,Martineau2004}.  In this
section we discuss in particular how our results can be used to
improve the design of new arrays of millimetre detectors for future
space missions such as SPICA-SAFARI
\citep{Goicoechea2011,Spinoglio2011}, PIXIE \citep{Kogut2011}, and
CORE \citep{CORE2011}.

The goal of  improving the noise-equivalent power (NEP) of new
experiments by at least one order of magnitude (from
$10^{-17}$\,W\,Hz\,$^{-1/2}$ to $10^{-18}$\,W\,Hz\,$^{-1/2}$) can be
achieved by increasing the focal-plane coverage, using thousands of
background-limited instrument performance (BLIP) contiguous
pixels.\footnote{The sensitivity of bolometers is fundamentally
  limited by statistical fluctuations of the radiation power
  from the observed source and the thermal emission from the cryogenic stages
  of the instrument.} Each pixel of these arrays must be micromachined
starting from a common substrate. 

Two new generations of detectors other than classical high-impedance
bolometers are in competition in the development of future millimetre
space missions. We summarize their principal characteristics and discuss
how  to minimize the effect of particle impacts.

\paragraph{Transition-edge sensors (TES)} are thermal detectors that
use a superconductor thermometer biased at the superconductor
transition ($R_{\rm mean} \sim 10$ m$\Omega$)
\citep[e.g.,][]{Piat2002,Hubmayr2009}. The only conceptual difference
between the HFI high-impedance bolometers and TES bolometers is the
sensitive element used to measure the change in temperature, which is
a superconducting element in TES. Direct CRs impacting TES pixels will
have the same effect as in the high-impedance bolometers. However, because of the common silicon wafer, the ballistic phonons
and thermal heat transport from the wafer to the thermometer will
affect several detectors at the same time in TES arrays. This effect could be
reduced, for example, by increasing the heat capacity of the substrate
and/or directing the ballistic phonon leak toward the housing and not to the pixels.

\paragraph{Kinetic inductance detectors (KIDs)} are superconducting resonant
  elements electromagnetically coupled to a common read-out
  transmission line. The absorption of the incoming photons causes a
  change of the resonance frequency through the breaking of Cooper
  pairs. The main advantages of KIDs over TES arrays
  are the straightforward frequency multiplexing readout and a
  faster time response (always $<1$\,ms). Another conceptual
  difference is that KIDs are insensitive to a change of base
  temperature provided that $T_{\rm base} \ll T_{\rm c}$ (where $T_{\rm c}$ is the
  superconductivity critical temperature). Considering an absorber of
  the same geometry as TES bolometers (suspended grid structure), the
  effect of particles in KIDs will be reduced by a factor between 10
  and 100 due to the (still poorly understood) capability of ballistic
  phonons to break Cooper pairs. Only phonons with energy exceeding
  the superconducting gap can break pairs and produce a measurable
  signal. The presence of such energetic ballistic phonons could be
  attenuated, as for the TES detectors, by increasing the leak
  of energy toward the housing. Work is in progress on this subject
 \citep{Swenson2010,Moore2012,Cruciani2012}.

\section{Conclusion}

We have described several ground-based tests that
lead to a physical interpretation of the impact of CR collisions on
the \Planck\ HFI bolometers in flight. An analytical model was developed 
to validate the physical interpretation with both ground-based and
in-flight data. Our results are the following:
\begin{itemize}

\item The dominant component of the in-flight long glitches is
 produced by the impact of CRs on the silicon
  wafer that supports the absorber and the NTD thermometer. The heat propagates 
  to the NTD thermometer by ballistic phonons followed by a contribution of thermal diffusion.

\item Short glitches are produced by direct interaction of a particle
  with either the absorber grid or the NTD-Ge thermistor.  In the case
  of PSB detectors, short glitches may also be the result of secondary
  electrons from a short glitch induced by a CR in the other PSB
  detector of the pair.

\item Slow glitches have not been reproduced in the ground-based
  tests. Their physical origin is consistent with the hypothesis that
  slow glitches result from the impact of CRs on the feed-through of
  the PSBa bolometers, but further work is needed to clearly
  identify their origin.

\item The influence of CR/detector glitches on scientific data quality needs to be included as  
  a design constraint for future-generation detector arrays for space applications.
  This needs to be included in
  parallel with all of the other characteristics such as detector NEP and time
  response. In particular, optical beam-response tests needs to be planned to study
  irradiation on complete detector arrays (e.g., pixels, substrate, and
  housing).  Starting from the same design conditions (for example,
  size and shape of the absorber, substrate and thermal contacts),
  non thermal detectors such as KIDs may present a conceptual advantage.

\end{itemize}

\bibliographystyle{aa}
\bibliography{cata2}

\end{document}